# A Robust and Novel Linear Fiber Laser Mode-locked by Nonlinear Polarization Evolution in All-polarization-maintaining Fibers

Xuanyi Liu, Qian Li, Denghui Pan, Feng Ye, Boris A. Malomed, and H. Y. Fu

*Abstract*—We demonstrate a novel, robust and compact fiber laser mode-locked by nonlinear polarization evolution (NPE) in polarization-maintaining (PM) fibers. The reflectivity of the artificial saturable absorber (SA) is analyzed to explain the mode-locking mechanism in the laser cavity. Experimentally, three linear laser schemes that feature repetition rates 94 MHz, 124 MHz and 133 MHz are systematically investigated. When the pump power is 1100 mW, the 124-MHz laser cavity delivers highly stable pulses with a single-pulse energy of 0.92 nJ. After the compression, the pulse duration obtained from the 124-MHz fiber laser is 250 fs, while the corresponding transform-limited pulse duration is 124 fs. The highest fundamental repetition rate that could be achieved in our experiment is 133 MHz, as mentioned above. The noise characterization has been performed with different cavity lengths and therefore different net-cavity dispersion. The 68-fs timing jitter and the 0.01% relative intensity noise (RIN) of the 133-MHz fiber laser have been realized integrated from 1 kHz to 10 MHz. Furthermore, the root-mean-square (RMS) power fluctuation is 0.35% in 2 hours, which implies superior stability of the output power. Thus, this linear fiber oscillator provides a competitive low-noise light source for optical applications appropriate for complex environments.

## I. INTRODUCTION

THE increasing number of industrial and scientific applications has caused great demand for lasers with superior optical performance and environmental stability. As competitive alternatives to solid-state mode-locked lasers, passively mode-locked fiber lasers offer excellent platforms for ultrashort pulse generation due to low manufacturing cost, compact structure, high conversion efficiency and good beam quality [1], [2]. Crucially important issues which need to be addressed in the design of high-performance mode-locked fiber lasers are the selection of the mode-locking mechanism, laser configuration, and appropriate fibers.

To produce high-quality ultrashort pulses, various mode-locking mechanisms have been developed. Real saturable absorbers (SAs), such as semiconductor saturable absorber mirror (SESAM) [3], or two-dimensional materials [4] in particular, graphene [5], are often used, enabling the mode-locked fiber lasers to acquire self-starting ability. However, the major limitations of real SA-based mode-locked fiber lasers are their low damage threshold and the fact that they provide the generation of relatively wide pulses. Artificial SAs may be based on the nonlinear polarization evolution (NPE) [6], [7], nonlinear optical loop mirror (NOLM) [8], and its extension in the form of a nonlinear amplifying loop mirror (NALM) [9], [10]. Among them, the NPE mode-locking technique has been intensively employed, as it provides fast saturation of the absorption, a high damage threshold, and a weak intrinsic noise. Combined with the cavity-dispersion engineering, NPE-based mode-locked fiber lasers can generate narrow pulses with high energy and a low noise level. By adjusting the group-velocity dispersion (GVD), cavity length and spectral-filtering bandwidth in an all-normal-GVD fiber laser based on NPE, the pulse width can be made as small as 31 fs, with the energy attaining 84 nJ [11]. It is relevant to mention that the fiber-laser scheme which does not include a specifically inserted bandpass filter can produce stable dissipative solitons as well, as the entire setup may generate a filtering effect [12]. Ma *et al*. have demonstrated the operation of an erbium-doped-fiber (EDF) laser, mode-locked by means of NPE, which delivers 37.4 fs pulses at a repetition rate of 225 MHz [13]. Further, Chen *et al*. have presented an NPE-based fiber oscillator with low jitter and low intensity noise [14]. It operates in the anomalous-GVD region, producing 167 fs soliton pulses at the repetition rate of 194 MHz. Nevertheless, the optical performance of NPE-based fiber lasers is subjected to environmental perturbations, such as temperature changes or mechanical vibrations, which hinder

This work was supported in part by Overseas Research Cooperation Fund of Tsinghua Shenzhen International Graduate School (HW2020006), in part by Shenzhen Science and Technology Innovation Commission (Project GJHZ20180411185015272), in part by Youth Science and Technology Innovation Talent of Guangdong Province (2019TQ05X227), and in part by Israel Science Foundation (grant No. 1286/17). *(Corresponding authors: Qian Li; H. Y. Fu.)*

Xuanyi Liu, Denghui Pan, and H. Y. Fu are with the Tsinghua Shenzhen International Graduate School and Tsinghua-Berkeley Shenzhen Institute, Tsinghua University, Shenzhen 518055, China (e-mail: hyfu@sz.tsinghua.edu.cn).

Feng Ye and Qian Li are with the School of Electronic and Computer Engineering, Peking University, Shenzhen 518055, China (e-mail: liqian@pkusz.edu.cn).

Boris A. Malomed are with the Department of Physical Electronics, School of Electrical Engineering, Faculty of Engineering, and Center for Light-Matter Interaction, Tel Aviv University, Tel Aviv 69978, Israel; and the Instituto de Alta Investigación, Universidad de Tarapacá, Casilla 7D, Arica, Chile (e-mail: malomed@tauex.tau.ac.il).

their practical applications.

Polarization-maintaining (PM) fibers can be introduced to overcome this drawback and make NPE fiber lasers more adaptable to complex environments. Two kinds of laser schemes including ring and linear fiber cavities have been proposed to achieve NPE mode-locking in PM fibers. One scheme involves pulse evolution in several segments of PM fibers spliced at specific angles, which assists mode-locking or acts as a mode-locker in the fiber laser. In 2016, Wang et al. have reported the operation of a ring NPE fiber laser with a repetition rate of 43.8 MHz. Cross-splicing method is employed to compensate the birefringence in PM fibers [15]. In 2017, Szczepanek et al. have demonstrated an all-PM 22.54-MHz fiber laser using three pieces of PM fibers, which are carefully calibrated and spliced. The pulses generated by such a fiber oscillator are compressible to 150 fs [16]. More recently, splicing three PM gain fibers and five standard PM ones have been demonstrated to provide the NPE mode-locking mechanism in PM fiber lasers [17], [18]. Zhou et al. have presented 2.9 nJ, 5.9 ps dissipative soliton pulses generation from a half-ring PM-NPE fiber laser, with a Faraday mirror (FM) utilized to eliminate the group-velocity mismatch (GVM) [19]. Impressively, Zhang et al. have reported a ring PM-NPE fiber laser, which contains free space and specially spliced PM fibers. This fiber oscillator delivers pulses at a high repetition rate of 111 MHz [20]. However, in the above-mentioned PM-NPE fiber lasers, the realization of the NPE mode-locking requires off-axis fusion splicing of PM fibers, which introduces additional loss in the fiber laser and increases the laser fabrication complexity. In linear laser schemes, an FM is placed at the end of the laser cavity. The function of the FM is to compensate the linear phase shift and GVM. In 2007, a linear all-PM fiber laser mode-locked by NPE has been demonstrated to emit 5.6 ps pulses at a repetition rate of 5.96 MHz [21]. By inserting a long PM fiber into the laser cavity, Boivinet et al. have presented a low-repetition-rate (948 kHz) linear fiber laser [22]. In 2020, Yu et al. have performed an experimental and numerical investigation of different pulses generated in an all-PM linear fiber laser [23]. Apart from the specially processed splicing angles between PM fibers, the NPE mode-locking in linear fiber lasers needs relatively long PM fibers to accumulate a sufficient nonlinear phase shift, which impairs the stability of the fiber laser. The inclusion of a gain fiber into the artificial SA has proven to be beneficial for achieving high repetition rates [24], and non-reciprocal phase shifters are widely employed in figure-nine fiber setups, mode-locked by NALM to build high-repetition-rate fiber lasers [25]. Still, it is challenging to design an all-PM linear fiber laser working at high repetition rates, because of the difficulty with designing a simple and compact laser configuration and manufacturing high-gain PM fiber.

In our previous work, we have implemented an NPE-based fiber laser containing PM-fiber and free-space portions. Our linear fiber laser achieves the highest repetition rate of 115 MHz and exhibits outstanding environmental stability [26]. However, a systematical investigation of the influence of different cavity lengths on its performance is missing, and the noise level of such kind of fiber lasers remains unknown.

In this paper, we aim to further optimize the laser configuration to improve the simplicity, compactness and robustness of the fiber oscillator. A novel linear PM fiber laser based on NPE is presented in details. By adjusting the laser-cavity length, optical characteristics of the generated pulses at different repetition rate (94 MHz, 124 MHz, 133 MHz) are systematically investigated and analyzed. Two key features of this linear laser scheme need to be emphasized. One is that the 124-MHz fiber laser delivers 0.92-nJ high-energy pulses with a spectral bandwidth of 28.8 nm. The pulses are compressible down to 250-fs duration. The other essential feature is 133 MHz which is the highest repetition rate among PM fiber lasers mode-locked by NPE, to the best of our knowledge. The 133-MHz fiber laser exhibits a low timing jitter of 68 fs and the relative intensity noise (RIN) of 0.01%, integrated from 1 kHz to 10 MHz. Tests of the average output power and spectral stability are reported, showing good practicability of this type of fiber laser.

## II. EXPERIMENTAL SETUP AND THEORETICAL ANALYSIS

Figure 1 shows the schematic diagram of the linear all-PM mode-locked fiber laser, which comprises two sections of free space and one segment of PM fiber. In the PM fiber portion, a 40-cm PM-EDF (Er80-8/125-PM, LIEKKI) is employed as the gain medium, which has a peak core absorption of ~80 dB/m at 1530 nm. The rest is the standard PM single-mode fiber. A commercial 976 nm laser diode (LD) is employed as the pump light source. It is composed of two single-mode diodes to provide a maximum pump power of 1.5 W. A key element of the design is that the wavelength-division multiplexer (WDM), the collimator, and the PM gain fiber are integrated into a single optical component (WDM-collimator), which greatly simplifies the design of the fiber laser. The detailed structure of the WDM-collimator is presented beneath the laser's scheme. The G-lens and filter play the role of the collimation and isolation, respectively. The pump light entering from the reflection port is coupled into the common port formed by the gain fiber. Two sections of free space are placed on both sides of the PM fiber. To increase the coupling efficiency and reduce the alignment loss, the lengths of the free-space sections are shrunk to 2 cm and 3.2 cm, respectively. As shown in Fig.1, for 2-cm free space part, a Faraday rotator (FR) and a $\lambda/8$-wave plate (EWP) are inserted into the laser cavity between a highly reflective mirror (M1) and a collimator (Col). The 3.2-cm free-space part contains a WDM-collimator, a half-wave plate (HWP), a polarization beam splitter (PBS), and another highly reflective mirror (M2).

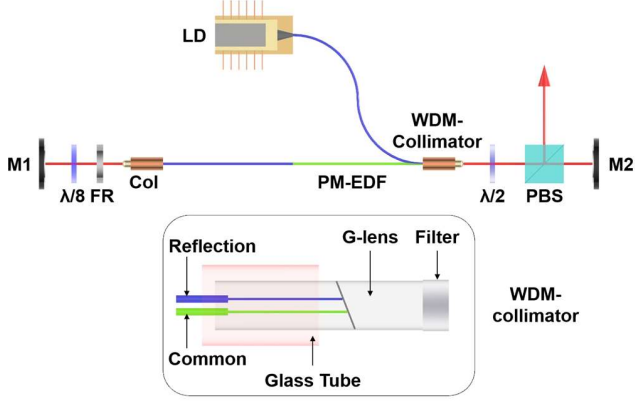

Fig. 1. Schemes of the all-PM mode-locked fiber laser and WDM-collimator.

In a NALM fiber laser, the mode-locking behavior depends on the nonlinear phase-shift difference between counter-propagating pulses in a fiber loop. An additional non-reciprocal phase shifter, together with the intensity-dependent nonlinear phase shift, enables easy self-starting operation [27]. Different from the NALM fiber laser, there is no fiber loop in this linear fiber setup. However, the fast and slow axes of the PM fibers play the same roles as opposite optical paths in a fiber loop. To investigate how the pulse evolves in the laser cavity, and understand the mode-locking mechanism, the intra-cavity roundtrip reflectivity needs to be theoretically analyzed as a function of the nonlinear phase shift. The linear conversion of the polarization state of light by each optical element in the laser can be expressed by the Jones matrices [28], as listed in table 1.

TABLE I
JONES MATRICES FOR OPTICAL ELEMENTS

| Optical elements | Jones matrix |
|---|---|
| Half-wave plate | $M_{HW}(\theta) = \begin{pmatrix} \cos\theta & -\sin\theta \\ \sin\theta & \cos\theta \end{pmatrix} \begin{pmatrix} e^{\frac{i\pi}{2}} & 0 \\ 0 & e^{\frac{i\pi}{2}} \end{pmatrix} \begin{pmatrix} \cos\theta & \sin\theta \\ -\sin\theta & \cos\theta \end{pmatrix}$ |
| λ/8-wave plate | $M_{EW}(\theta) = \begin{pmatrix} \cos\theta & -\sin\theta \\ \sin\theta & \cos\theta \end{pmatrix} \begin{pmatrix} e^{-\frac{i\pi}{8}} & 0 \\ 0 & e^{\frac{i\pi}{8}} \end{pmatrix} \begin{pmatrix} \cos\theta & \sin\theta \\ -\sin\theta & \cos\theta \end{pmatrix}$ |
| Faraday rotator (45°) | $M_{FR} = \begin{pmatrix} \frac{\sqrt{2}}{2} & \frac{\sqrt{2}}{2} \\ -\frac{\sqrt{2}}{2} & \frac{\sqrt{2}}{2} \end{pmatrix}$ |
| Polarization beam splitter | $M_{PBS,trans} = \begin{pmatrix} 1 & 0 \\ 0 & 0 \end{pmatrix}$ |
| Mirror | $M_M = \begin{pmatrix} -1 & 0 \\ 0 & -1 \end{pmatrix}$ |
| Nonlinear phase bias | $M_{NLP1} = \begin{pmatrix} e^{i\Delta\varphi_{nl}/2} & 0 \\ 0 & 1 \end{pmatrix}$, $M_{NLP2} = \begin{pmatrix} 1 & 0 \\ 0 & e^{i\Delta\varphi_{nl}/2} \end{pmatrix}$ |

The polarization state of the incident pulse transmitted by the PBS can be expressed as $E_x = (1, 0)$. The nonlinear phase shift, $\Delta\varphi_{nl}$, actually represents the nonlinear phase-shift difference between the two pulses with orthogonal polarizations, when they propagate through the PM fiber. After double-passing the PM fibers and free-space optical components, the electric field $E_{trans} = (E_x, E_y)$ can be calculated as

$$E_{trans} = M_{PBS,trans} M_{HW}(\theta_h) M_{NPL1} M_{FR} M_{EW}(\theta_e) \\ M_M M_{EW}(\theta_e) M_{FR} M_{NPL2} M_{HW}(\theta_h) E_x, \quad (1)$$

where $\theta_h$ and $\theta_e$ represent the rotation angle of the HWP and the EWP, respectively. The roundtrip reflectivity function is then derived as

$$R = |E_x|^2. \quad (2)$$

In this linear fiber setup, a change of $\theta_h$ and $\theta_e$ can greatly affect the pulse evolution, hence also the roundtrip reflectivity of the artificial SA. Two specific sets of wave plate angles are analyzed to investigate the relationship between the reflectivity and nonlinear phase shift. Figure 2(a) depicts the situation when $\theta_h$ is fixed at 22.5º, meaning that the incident linearly polarized pulse transmitted by the PBS is divided by HWP in orthogonal pulses with a certain intensity ratio. When $\theta_e$ changes from 0.0º to 45.0º, the reflectivity function shifts to left, giving rise to different phase biases with an increased modulation depth. The resulting steeper slope at $\Delta\varphi_{nl} = 0$ allows more high-intensity pulses to be generated in the laser cavity, enhancing the mode-locking ability. In the case of $\theta_e = 45.0º$, a π/2 nonlinear phase shift with 100% modulation depth can be observed, where the orthogonal pulses are rotated by 45.0º by the FR, and coincide with the fast and slow axes of the EWP if the rotation by 45.0º is applied to the EWP. Under such a condition, the combination of the FR and EWP gives a maximum nonlinear phase bias, and the pulses with orthogonal polarizations can achieve complete interference. In Fig. 2(b), changes in the reflectivity function with fixed $\theta_e = 45.0º$ are demonstrated. When $\theta_h$ is reduced from 22.5º to 0.0º, each curve acquires a nonlinear phase shift of π/2 with a decreased modulation depth. The variation of the modulation depth is attributed to a change in the intensity distribution ratio provided by the HWP. No modulation depth can be observed when $\theta_h$ is adjusted to 0.0º. In this case, the incident pulse is projected exactly onto the fast or slow axis of the HWP, with no pulse reflected back.

The above analysis makes it clear that the pulse evolution in this setup is totally different from the result we reported recently [26]. Firstly, the HWP is responsible for decomposing the pulse into orthogonal components. After passing the HWP, the two pulses are transmitted being aligned with the fast and slow axes of the PM fiber. Then, the FR offers a 45-degree rotation and the rotation of the EWP (45.0º) makes the orthogonal pulses exactly corresponding to the fast and slow axes of the EWP. As a result, an additional non-reciprocal phase bias of π/2 can be provided, permitting the laser to work at low intensities and improving the self-starting ability. Finally, the pulses with the orthogonal polarizations are reflected by the end-mirror and propagate through all the free-space optical elements and the PM fiber again. The PBS is placed where the orthogonal pulses interfere with each other and create linearly polarized pulses emitted from the laser cavity. This process, named artificial SA, supports passively mode-locking in the fiber laser.

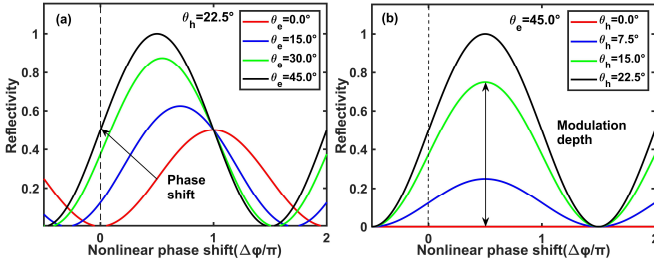

Fig. 2. The theoretical roundtrip reflectivity of the artificial SA as a function of the nonlinear phase shift. (a) At $\theta_h = 22.5°$, adjusting $\theta_e$ changes both the phase shift and modulation depth; (b) At $\theta_e = 45.0°$, the variation of $\theta_h$ only changes the modulation depth with a fixed $\pi/2$ phase shift.

## III. EXPERIMENTAL RESULTS AND DISCUSSIONS

### A. Mode-locked pulse characteristics

Three linear laser schemes operating at different repetition rates (94 MHz, 124 MHz, and 133 MHz) have been built by choosing the length of the non-gain fiber (0.66 m, 0.39 m, and 0.34 m). To produce stable mode-locked pulses, the pump power of the LD and rotation angles of the HWP and EWP need to be precisely adjusted. The mode-locking operation can be separated into the following four steps. Firstly, the pump power of the LD is increased to sufficiently large values. Secondly, the HWP and EWP are appropriately rotated to acquire a nonlinear phase bias, allowing more high-intensity spikes to occur in the laser cavity. Thirdly, once the laser switches from the continuous-wave (CW) operation to the pulsed regime, gradually decreasing the pump power can eliminate the CW component and achieve clean mode-locking. Finally, angles of the wave plates are slightly tuned to optimize the spectral and temporal characteristics without affecting the mode-locking operation. Once the wave plates are placed in appropriately rotated positions, directly ramping up the pump power beyond the mode-locking threshold can achieve self-starting, and the laser does not require any additional tuning.

Figures 3(a)-(f) depict spectral features of the mode-locked pulses on the logarithmic and linear scales, when the laser is adapted to different cavity lengths, hence different repetition rates. The output spectrum has been measured by an optical spectrum analyzer (AQ6370D, YOKOGAWA). The spectra show typical soliton mode-locking characteristics, with observable Kelly sidebands distributed on both sides. The asymmetry of the sidebands may be induced by the third-order dispersion (TOD) [29]. The logarithmic spectrum is characterized by two main peaks and rough sides, which is mainly attributed to the interference of the orthogonally polarized pulses at the PBS. Since the generated dissipative solitons coexist with complex and dense sidebands, soliton distillation offers an effective method to obtain pure solitons from the resonant CW background [30]. Compared to the regimes of the operation at 94 MHz and 133 MHz repetition rates, the regime corresponding to 124 MHz exhibits a broader spectral full width at half-maximum (FWHM), viz., 28.8 nm. In this regime, the fiber laser works at a higher pump power and generates mode-locked pulses with a higher single-pulse energy. The self-phase modulation (SPM) in the cavity is enhanced, resulting in the broadest spectrum. Besides, the total length of the PM fiber portion is 0.79 m, including a 0.4-m gain-fiber segment and a 0.39-m non-gain one. The short cavity length makes the net cavity GVD closer to zero, leading to a larger spectral bandwidth. The average output power is recorded by a power meter (S148C, Thorlabs). When the pump power of the LD is 1100 mW, the 124-MHz fiber laser emits mode-locked pulses with the average output power of 114 mW, corresponding to a single-pulse energy of 0.92 nJ.

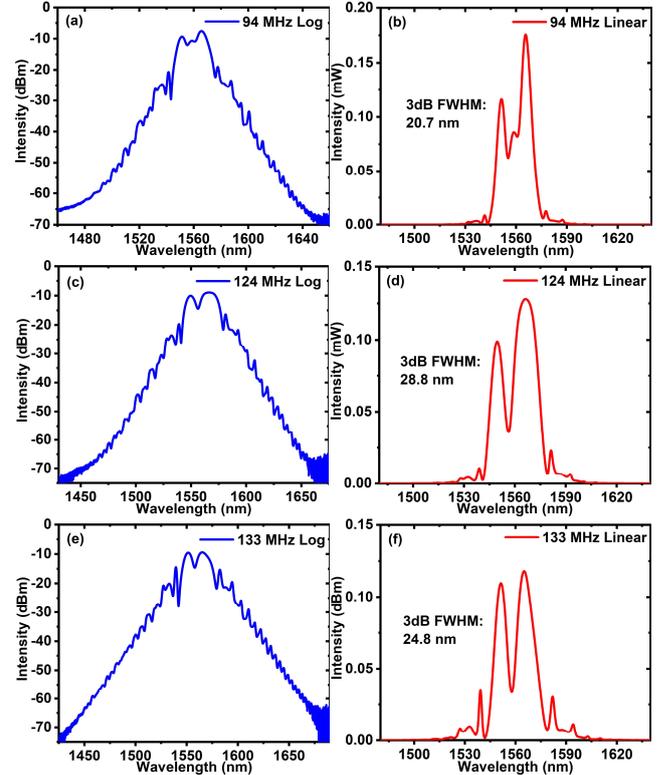

Fig. 3. Optical spectra measured on the logarithmic and linear scales (blue and red lines, respectively) at different repetition rates.

To investigate the mode-locking properties of the fiber laser, radio-frequency (RF) spectra and pulse trains generated by these three laser schemes have been characterized. The RF spectra have been measured using a RF signal analyzer (N9030B, Agilent) with a bandwidth from 3 to 50 GHz, as presented in Fig. 4(a)-(c). The peak-to-background ratio of the fundamental repetition rate is better than 80 dB for each of the RF spectra centered at 94 MHz, 124 MHz and 133 MHz. No other RF spectral components are observed on either side of the peak. The wideband RF spectra are presented in insets of Fig. 4(a)-(c), which confirms the single-pulse operation. There is no sinusoidal modulation or multi-pulsing effects, indicating good pulse-to-pulse stability. The pulse train is measured by a 3 GHz bandwidth InGaAs photodetector (PD) and a real-time oscilloscope (DSO-X 6004A, Keysight). Figures 4(d)-(f) reveal the pulse sequences at equal time intervals of 10.6 ns, 8.1 ns and 7.5 ns, which correspond, respectively, to fundamental repetition rates of 94 MHz, 124 MHz and 133 MHz. The flat upper and lower interfaces of the pulse train verify excellent temporal performance of the output pulses.

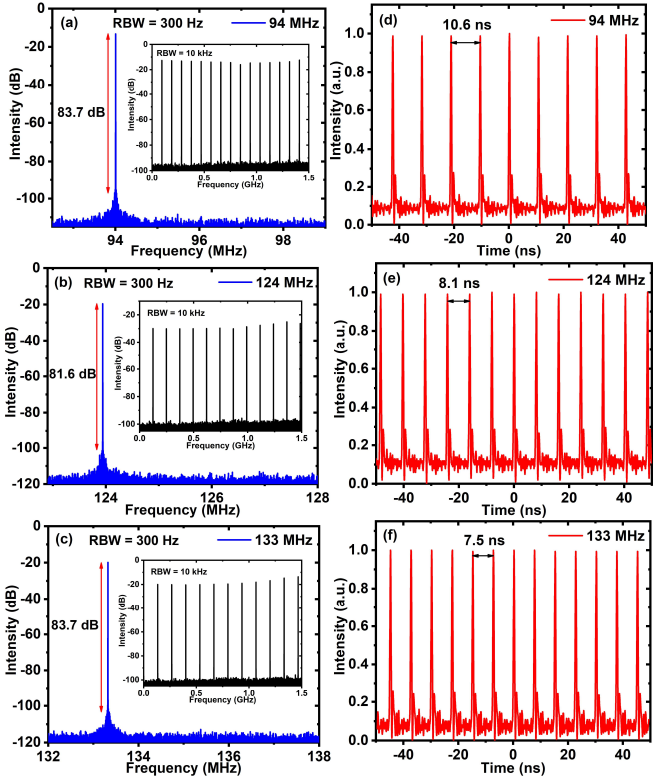

Fig. 4. The RF spectra (a)-(c) and mode-locked pulse trains (d)-(f) of the linear PM fiber laser at different repetition rates.

The autocorrelation traces of the mode-locked pulses are detected by a commercial intensity autocorrelator (FR-103XL/IR/FA, Femtochrome). As shown in Figs. 5(a)-(c), for mode-locked pulses directly generated from the laser cavity at repetition rates of 94 MHz, 124 MHz and 133 MHz, the corresponding pulse widths are 0.74 ps, 0.91 ps and 1.35 ps, respectively. The measured pulse durations, calculated on the basis of the spectral FWHM, are far from the Fourier transform limit (173 fs, 124 fs and 144 fs). The generated pulses are highly chirped because all PM fibers and free-space optical elements exhibit anomalous GVD. The intra-cavity dispersion of the three setups considered here is estimated to be -0.023 ps$^2$, -0.017 ps$^2$ and -0.016 ps$^2$, respectively. Due to the wide spectral bandwidth, the external pulse compression for the 124-MHz pulses has been carried out using a segment of a dispersion compensation fiber (DCF), whose GVD is -160 ps/nm/km at the wavelength around 1550 nm. Autocorrelation traces of the compressed pulses are plotted in Fig. 5(d). The pulses generated by the 124-MHz fiber laser can be compressed down to 250 fs, which is almost twice the Fourier-transform limit (124 fs). This fact has two aspects. Firstly, the residual GVD is not completely compensated by the DCF with its actual length. Secondly, the TOD and nonlinear chirp also restrict further compression of the pulse. The TOD is mainly contributed by the PM fibers, used to construct the laser cavity, and by the DCF employed for the pulse compression. When the ultrashort pulses double-pass the PM fibers and single-pass the DCF, the total TOD is calculated to be -0.000231 ps$^3$ for the 124-MHz fiber laser.

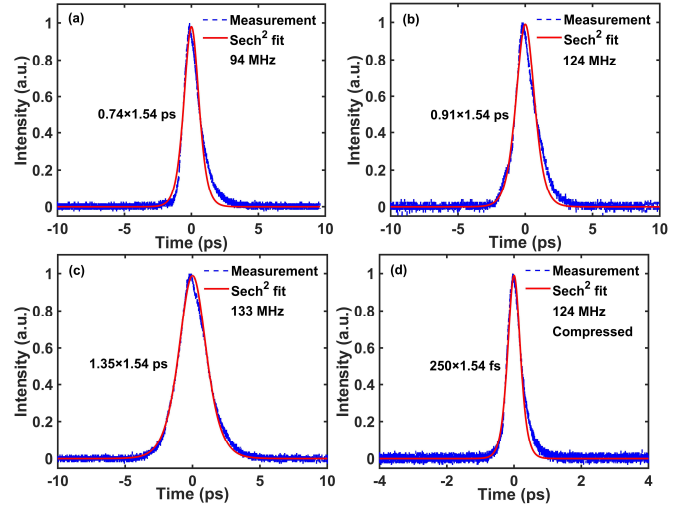

Fig. 5. Autocorrelation traces of the ultrashort pulses (a)-(c) produced by the linear PM fiber laser at different repetition rates; (d) Compressed pulse duration for the 124-MHz fiber laser.

Table 2 provides detailed comparison of the performance parameters of the laser setups operating at different repetition rate. To maintain sufficient single-pulse energy in the high-repetition-rate operation under the low coupling efficiency of the WDM-collimator, a high pump power is required to self-start mode-locking. However, the mode-locked spectrum includes the CW component, and the single-pulse operation is observed at relatively lower pump powers. In Table 2, all optical parameters are referred to the single-pulse operation. When the length of the PM gain fiber is fixed to be 0.4 m, it is found that the 124-MHz fiber laser can emit pulses with a high energy of 0.92 nJ. The 133-MHz fiber laser operates at a relatively low pump power and generates 0.78-nJ pulses, which is caused by the effect of the splicing loss in the experiment. The single-pulse energy of the generated pulse trains with different repetition rates exceeds the soliton's energy. The pulse with such a high energy does not split, which is mainly due to the compact laser structure and the unique mode-locking mechanism. It is difficult for the fiber laser to achieve mode-locking when the non-gain fiber is made still shorter, with the aim to increase the repetition rate, limited by the maximum pump power of the LD. Thus, 133 MHz is the highest repetition rate of the fiber laser which could be achieved in our experiment. Obtaining still higher repetition rate requires injecting higher pump power into the laser cavity or using high-gain PM fibers.

TABLE II
OPTICAL PERFORMANCE OF THE LINEAR PM FIBER LASER AT DIFFERENT RATES

| $F_{rep}$ (MHz) | $L_{SMF}$ (m) | $L_{EDF}$ (m) | $P_{pump}$ (mW) | $\Delta\lambda$ (nm) |
|---|---|---|---|---|
| 94 | 0.66 | 0.4 | 460 | 20.7 |
| 124 | 0.39 | 0.4 | 1100 | 28.8 |
| 133 | 0.34 | 0.4 | 960 | 24.8 |
| $F_{rep}$ (MHz) | $P_{av}$ (mW) | $E_p$ (nJ) | $\tau_p$ (ps) | $\tau_{tl}$ (fs) |
| 94 | 45.5 | 0.48 | 0.74 | 173 |
| 124 | 114.0 | 0.92 | 0.91 | 124 |
| 133 | 103.8 | 0.78 | 1.35 | 144 |

$F_{rep}$, the repetition rate; $L_{SMF}$, length of the non-gain fiber; $L_{EDF}$, length of the gain fiber; $P_{pump}$, pump power of the laser diode; $\Delta\lambda$, spectral full width at half-maximum; $P_{av}$, average output power; $E_p$, single-pulse energy; $\tau_p$, pulse duration; $\tau_{tl}$, transform-limited pulse duration

## B. Amplitude and phase noise characteristics

The noise characteristic is an important indicator of the environmental impact on laser performance (see, e.g., Ref. [12]). The phase and amplitude noise has been measured under the free-running condition of the linear PM fiber laser. A few milliwatts of output mode-locked pulses are coupled into the photodetector. The optical signal is converted into the electrical signal, then characterized by a phase noise analyzer (FSWP8, Rohde & Schwarz). Figures 6(a)-(b) depict the measured phase noise and RIN curves in the range of offset frequencies from 10 Hz to 10 MHz. The phase noise spectra for fiber lasers operating at the three above-mentioned values of the repetition rate show a common downward trend. The phase noise level decreases as the repetition rate of the fiber laser increases. This can be attributed to the fact that the intra-cavity GVD of the 133-MHz fiber laser is closer to zero. The RIN curves for each repetition rate are plotted in Fig. 6(b). The setup with rate 124 MHz operates at higher pump power (1100 mW), thus exhibiting a higher RIN level compared to other two values of the repetition rate. Because the RIN of the mode-locked fiber laser is dominated by the pump-laser's intensity noise [31]. The RIN spectra corresponding to repetition rates 94 MHz and 133 MHz are almost identical, which is the result of the combined effect of intra-cavity GVD and pump power. In addition, multiple noise spikes can be observed in the frequency range from 10 Hz to 1 kHz, which is mainly caused by acoustic noise or mechanical vibrations in the laboratory environment [32]. Placing the fiber laser in a sheltering box can suppress external environmental noise.

Compared to the setups corresponding to the other two values of the repetition rate, better noise characteristics are achieved by the 133-MHz fiber laser, whose noise spectra and integrated noise are presented in Figs. 6(c)-(d) separately. In the frequency range from 10 Hz to 100 kHz, the phase noise drops from -60 dBc/Hz to -160 dBc/Hz. The noise curve gradually flattens and attains the lowest value of -168 dBc/Hz as the frequency offset grows from 100 kHz to 10 MHz. The RIN is between -170 dBc/Hz and -110 dBc/Hz if the noise spikes are ignored. The timing jitter of 68 fs and the integrated RIN of 0.01% are featured by the 133-MHz fiber laser. The excellent noise performance makes this fiber laser competitive in comparison to other pulsed light sources [33], [34].

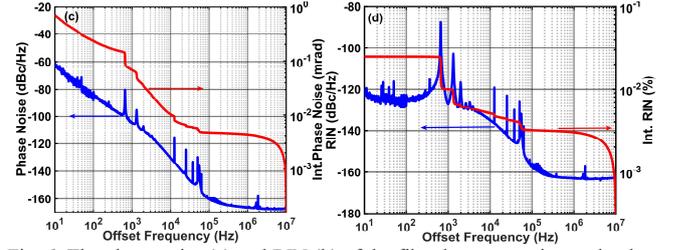
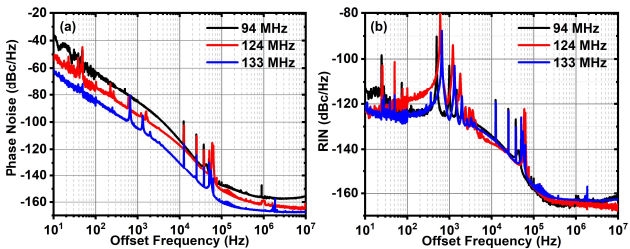

Fig. 6. The phase noise (a) and RIN (b) of the fiber laser operating at the three repetition rates; (c) The phase noise and integrated phase noise of the 133-MHz fiber laser; (d) The RIN and integrated RIN of the same setup.

## C. Stability characteristics

The practical use of the fiber laser outside the laboratory depends on whether the output pulse characteristics of the mode-locked pulses can maintain long-term stability [8]. We performed stability test for the 124-MHz fiber oscillator by exposing the laser to the open air under room temperature without any sheltering box. As shown in Fig. 7(a), an average output power of 114.0 mW with a power fluctuation of ~0.4 mW indicates a 0.35% root mean square (RMS). Benefiting from the simple and compact configuration, shaking the PM fiber or tapping the optical platform will not interrupt the mode-locking operation and the output characteristics. To check the spectral stability, the optical spectrum is monitored every 15 minutes in the course of 2 hours, as shown in Fig. 7(b). The shape and bandwidth of the spectrum remain unchanged, indicating the robustness of the mode-locking.

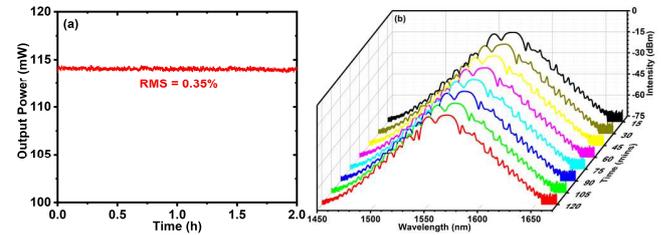

Fig. 7. (a) The average output power stability test; (b) The spectral stability test conducted in the course of two hours.

## IV. CONCLUSIONS

We have presented a novel and simple laser configuration for generating environmentally stable pulses. The theoretical analysis demonstrates that the $\pi/2$ non-reciprocal phase bias, together with power-dependent nonlinear phase shift, acts as an artificial SA (saturable absorber) in this NPE (nonlinear polarization evolution)-based fiber laser. Experimentally, the setups with different repetition rates (94 MHz, 124 MHz and 133 MHz) have been systematically investigated. Especially, 0.92-nJ energy pulses with the spectral FWHM of 28.8 nm are produced by the 124-MHz fiber laser. The repetition rate of 133 MHz, to the best of our knowledge, is the highest level ever reported for a PM fiber laser mode-locked by NPE. The performance of the fiber laser in the presence of the phase and amplitude noise has also been characterized. The timing jitter of 68 fs and RIN of 0.01%, integrated from 1 kHz to 10 MHz, have been achieved by the 133-MHz fiber laser. The average power fluctuation and spectral stability have been monitored in the course of two hours, indicating excellent stability of the output pulses for the fiber laser. We believe that this NPE-based linear fiber laser is a promising low-noise stable pulsed seed source for numerous practical applications.

**Xuanyi Liu** (Student Member, IEEE, Student Member, OSA) received the B.S. degree in the College of Electronic Science and Engineering from Jilin University, Changchun, China, in 2017, the Master degree in the School of Electronic and Computer Engineering, Peking University. He is currently pursuing the PhD degree with the Tsinghua Shenzhen International Graduate School and Tsinghua-Berkeley Shenzhen Institute, Tsinghua University.

**Qian Li** (Member, IEEE, Senior Member, OSA) received the Bachelor of Science degree from Zhejiang University, Hangzhou, China, in 2003, the Master of Science degree from the Royal Institute of Technology (KTH), Stockholm, Sweden, in 2005, and the Ph.D. degree from the Hong Kong Polytechnic University, Hong Kong, in 2009. After graduation she was a Visiting Scholar at the University of Washington, Seattle and Postdoctoral

Fellow at the Hong Kong Polytechnic University. In 2012 she joined School of Electronic and Computer Engineering (ECE) in Peking University as an Assistant professor. Since 2013 she is Associate Professor at ECE. Her research interests include nonlinear optics, ultrafast optics and integrated optics. Dr. Li is members of Institute of Electrical and Electronics Engineers (IEEE) and senior member of the Optical Society of America (OSA). From March 2017 to April 2019, she is Vice Chair of IEEE ED/SSC Beijing Section (Shenzhen) Chapter and Chair for EDS. From 2015 she is an advisor of OSA Student Chapter in Peking University Shenzhen Graduate School. From 2019 she is an advisor of Peking University Shenzhen Graduate School IEEE Photonics Society Student Branch Chapter.

**Denghui Pan** (Student Member, IEEE, Student Member, OSA) received the B.S. degree in the College of Chemical Engineering from Zhejiang University of Technology, Hangzhou, China, in 2017, He is currently pursuing the Master degree with the Tsinghua Shenzhen International Graduate School and Tsinghua-Berkeley Shenzhen Institute, Tsinghua University.

**Feng Ye** (Student Member, IEEE, Student Member, OSA) received the B.S. degree in photoelectric information science and engineering from South China Normal University, Guangzhou, China, in 2019. He is currently pursuing the Master of Science degree with the School of Electronic and Computer Engineering, Peking University.

**Boris A. Malomed** (Senior Member, OSA) received the Ph.D. degree from the Moscow Physico-Technical Institute, Russia, in 1981, and the Doctor's of Science degree (habilitation) from the Institute for Theoretical Physics of the Academy of Sciences of Ukraine (Kiev) in 1989. He has been working at the Tel Aviv University (currently, as a Professor with chair "Optical Solitons") since 1991. Since 2011, he has also been collaborating, as a Consultant, with the Institute of Photonic Sciences (ICFO, Barcelona, Spain). He was a divisional associate editor of Physical Review Letters (responsible for the area of "laser physics") in 2009–2015. Currently he is an editor of Phys. Lett. A and Chaos, Solitons & Fractals, and an editorial board member of J. Optics. His research interests include the fields of nonlinear optics, Bose–Einstein condensates and matter waves, pattern formation in nonlinear dissipative media, dynamics of nonlinear lattices, etc. His current h-index is 78 (as per Web of Science and Scopus) and 91 (as per Google Scholar)

**H. Y. Fu** (Senior Member, IEEE, Life Member, OSA) is currently an associate professor of Tsinghua Shenzhen International Graduate School (SIGS) and Tsinghua-Berkeley Shenzhen Institute, Tsinghua University. He received the B.S. degree in electronic and information engineering from Zhejiang University, Hangzhou, China, and the M.S. degree in electrical engineering with specialty in photonics from Royal Institute of Technology, Stockholm, Sweden, and the Ph.D. degree from the Department of Electrical Engineering from Hong Kong Polytechnic University. His research interests include integrated photonics and its related applications, fiber optical communications, fiber optical sensing technologies. From 2005 to 2010, he was a research assistant and then research associate with Photonic Research Center, the Hong Kong Polytechnic University. From 2010 to April 2017, he was a founding member and leading the advanced fiber optic communications research at Central Research Institute, Huawei. He was a project manager of All-Optical Networks (AON), which was evolved to a company-wide flagship research project that covers whole aspects of next generation optical communication technologies to guarantee Huawei's leading position. He was also a representative for Huawei at several industry/academic standards/forums. He was an active contributor at IEEE 802.3 Ethernet and Optical Internetworking Forum (OIF) where he was an OIF Speaker from 2012 to 2013. Dr. Fu is senior member of IEEE and life member of OSA, SPIE. From 2017, he is an advisor of OSA Student Chapter at TBSI, Tsinghua University. From 2020, he is advisor of IEEE Photonics Society Student Branch Chapter and SPIE Student Chapter at Tsinghua SIGS. He has authored/coauthored more than 180 journal or conference papers, 1 book chapter, over 50 grant/pending China /Europe/Japan/ US patents.